\documentclass[prd,aps,showpacs,nofootinbib,eqsecnum,onecolumn,groupedaddress,amssymb]{revtex4}
\usepackage{amsmath}
\usepackage{amssymb}
\usepackage{amsfonts}
\usepackage{graphicx,bm}
\usepackage{dcolumn}
\usepackage{color,amsxtra}
\usepackage{epsf}
\usepackage{enumerate}
\usepackage{hhline}
\usepackage{array}
\usepackage{tabularx}
\usepackage{subfigure}
\usepackage{fancyhdr}
\usepackage{mathrsfs}

\newcommand{\be}{\begin{equation}}
\newcommand{\ee}{\end{equation}}
\newcommand{\bea}{\begin{eqnarray}}
\newcommand{\eea}{\end{eqnarray}}
\newcommand{\beaa}{\begin{eqnarray*}}
\newcommand{\eeaa}{\end{eqnarray*}}

\newcommand{\Eqn}[1]{&\hspace{-0.2em}#1\hspace{-0.2em}&}

\def\be{\begin{equation}}
\def\ee{\end{equation}}
\def\bea{\begin{eqnarray}}
\def\eea{\end{eqnarray}}

\begin{document}

\title{\large Beyond-one-loop quantum gravity action yielding \\ both inflation and late-time acceleration}

\author{E.~Elizalde$^{1,2}$,
S.~D.~Odintsov$^{1, 2, 3}$,
L.~Sebastiani$^{4, 5}$,
and R.~Myrzakulov$^{4}$
}

\affiliation{ \\ \vspace*{2mm}
$^1$Consejo Superior de Investigaciones Cient\'{\i}ficas, ICE/CSIC-IEEC,
Campus UAB, Carrer de Can Magrans s/n, 08193
Bellaterra (Barcelona) Spain\\
$^2$International Laboratory for Theoretical Cosmology, Tomsk State University of Control Systems and Radioelectronics (TUSUR),
634050 Tomsk, and
  Tomsk State Pedagogical University, Tomsk, Russia\\
$^3$Instituci\'{o} Catalana de Recerca i Estudis Avan\c{c}ats
(ICREA), Barcelona, Spain\\
$^4$Department of General \& Theoretical Physics and Eurasian Center for
Theoretical Physics, Eurasian National University, Astana 010008, Kazakhstan\\
$^5$Dipartimento di Fisica, Universit\`a di Trento, Italy
}


\begin{abstract}
A unified description of early-time inflation with the current cosmic acceleration is achieved by
means of a new theory that uses a quadratic model of gravity, with the inclusion of an exponential
$F(R)$-gravity contribution for dark energy. High-curvature corrections of the theory come from
higher-derivative quantum gravity and yield an effective action that goes beyond the one-loop approximation.
It is shown that, in this theory, viable inflation emerges in a natural way, leading to a spectral
index and tensor-to-scalar ratio that are in perfect agreement with the most reliable Planck results.
At low energy, late-time accelerated expansion takes place. As exponential gravity, for dark energy, must be stabilized during the matter and radiation eras, we introduce a curing term in order to avoid nonphysical singularities in the effective equation of state parameter. The results of our analysis are confirmed by accurate numerical simulations, which show that our model does fit the most recent cosmological data for dark energy very precisely.
\end{abstract}


\maketitle

\def\thesection{\Roman{section}}
\def\theequation{\Roman{section}.\arabic{equation}}

\section{Introduction}

It is well accepted nowadays that the Universe underwent a period of strong and extremely quick accelerated expansion, namely the inflation stage, immediately after its origin (usually termed as the Big Bang singularity). From the very first proposal of the inflationary paradigm in 1981, by Guth~\cite{Guth} and Sato~\cite{Sato}, several attempts to describe this early-time acceleration have been carried out (see Refs.~\cite{infrev}, for some reviews).

Moreover, cosmological data~\cite{Planck} clearly show that the Universe is experiencing now a new phase of accelerated expansion, which can be explained either in terms of the existence of a dark energy fluid~\cite{Li:2011sd, Kunz:2012aw, Bamba:2012cp} or by modifying Einstein's gravity. In this respect, one of the most popular classes of modified gravity theories is $F(R)$-gravity. Here, the gravitational action is given by a function of the Ricci scalar only (for a review, see Refs.~\cite{reviewFr}). Many authors have investigated $F(R)$-gravity as an alternative for dark energy and its properties, showing that  theories of this class are able to fulfill the constrains imposed by local and cosmological tests~\cite{reviewFr, oldFR}.
In particular, exponential models of modified gravity as sound alternatives to dark energy have become quite popular in the last decade, since they represent a simple and natural way to mimic the cosmological constant term of the standard $\Lambda$CDM model at large curvature~\cite{nostriexp, altriexp, SuperBamba, oscillation1, oscillation2, oscillation3}.

As first suggested in Ref.~\cite{inlation+DE}, it may be interesting and natural---e.g., as the first step towards the construction of a more fundamental theory---to try to unify the early-time and late-time cosmological accelerations in one single model.
In this respect, it is worth noting that, at high curvature, when early-time inflation occurs, quantum gravity effects have to be incorporated to the theory. Starting from this crucial observation, we would like to present here high-curvature corrections to General Relativity (GR) under the form of a
higher-derivative quantum gravity model~\cite{B}. We thus enter the domain of quantum field theory, where basic renormalization group (RG) considerations lead to a RG improvement of the effective action.
Indeed, this technique has been successfully developed in quantum field theory in curved
space-time~\cite{El} and permits to construct an effective action which goes beyond the  one-loop approximation, because it renders it possible to sum over all leading log-terms of the theory. Since we are here interested in the Friedmann-Robertson-Walker (FRW) space-time solutions, we will be finally led to work with a higher-derivative multiplicatively renormalizable quantum gravity theory~\cite{Stelle, Fradkin} through the use of RG-improved techniques. A model of this kind was first discussed in Ref.~\cite{qg1} and subsequently extended to the case of $R^2$ with quantum electrodynamics in Ref.~\cite{qg2}. Here, we further extend the formulation in order to go beyond these results and obtain a unified description of a viable inflationary scenario with the Friedmann and dark energy Universe stages. To reproduce the dark energy sector we make use of an exponential model of $F(R)$-modified gravity, including an additional term to stabilize the theory during the radiation and the matter eras.
Our approach is phenomenological and we add the dark energy $F(R)$-function by hands. Of course, in that case we follow several conditions. First of all, we choose the exponential form of $F(R)$ so that it would be qualitatively similar to RG improved effective action under discussion.
Second, its choice is done in such a way that inflationary universe scenario which follows from our RG improved effective action is not modified by dark energy $F(R)$-term which gives non-essential impact to inflation. Third, our purpose is to formulate the unified description of the early-time inflation with dark energy. For that reason, our exponential $F(R)$-term is chosen so that to cancel some singularities (past-time oscillations) of the whole theory. Of course, the addition of such term to quadratic action makes the theory to be non-renormalizable.
In this sense, our complete action is classical modified gravity theory where some terms relevant for early-time inflation are inspired by quantum gravity considerations while the exponential $F(R)$-term is chosen so that it is negligible at the inflationary epoch.

The paper is organized as follows. In Section {\bf II} we discuss our model of a RG-improved effective action for higher-derivative renormalizable quantum gravity. The equations for the running coupling constants in front of the gravitational invariants are obtained, and explicit forms for these running coupling constants are derived. Section {\bf III} is devoted to the application of high-derivative quantum gravity to inflation. The field equations for FRW space-time are presented and the quasi-de Sitter (dS) solution describing the early-time acceleration is found. Furthermore, we show that this solution is unstable and that the model has a graceful exit from inflation, leading moreover to an amount of inflation (number of $e$-folds) that is large enough in order to get the necessary thermalization of the observable Universe. In Section {\bf IV} we derive the spectral index and the tensor-to-scalar ratio for cosmological perturbations. We show that these parameters are in agreement with the most recent analysis of the Planck satellite data. At the end of inflation, the model can be recast under the form of an $R^2$-correction to General Relativity, with a reheating mechanism able to convert the energy of inflation into the one corresponding to standard matter and radiation. In Section {\bf V} we introduce exponentially modified gravity for dark energy. We recover the de Sitter solution for the current cosmic acceleration, and show that this solution is a final attractor of the system. Modified gravity for dark energy needs a mechanism to avoid singularities during the radiation and matter eras, that is why in Section {\bf VI} we device a suitable logarithmic correction which stabilizes the theory at large curvature. In Section {\bf VII}, we provide a numerical simulation of the late-time acceleration occurring in our model. We should remark that the whole gravitational Lagrangian of the theory is here considered, which shows that the high-curvature corrections for inflation do not affect the dynamics of our model at late times. Actually, this model proves to be stable and to fit remarkably well the dark energy parameters coming from the latest analysis of Planck's data. Conclusions and final remarks are given in Section {\bf VIII}.

\section{RG-improved effective action for higher-derivative quantum gravity}

It is well-known that the Hilbert-Einstein action which classically describes the gravitational field gives rise to a non-renormalizable theory since, at high energy, the strong interaction is affected by divergences that cannot be canceled by a finite number of counter terms. This means that some ultraviolet completion is necessary at high energy scales. In this context, the development of a renormalizable quantum gravity theory is extremely interesting and can be also very useful in other areas of field theory. To wit, gravitation functions may act as a cut-off for interactions in elementary particle theory and it should be mentioned that a number of speculations are open in that sense.

For the pure gravitational action, renormalizability can be achieved by using quadratic terms on the Ricci scalar and Ricci tensor~\cite{B, Stelle, Fradkin, Hooft}, but the resulting theory is not ghost free and in the end some extension (eventually, non-local) is required~\cite{Stelleplus}.
The ghost problem in higher-derivative gravity is not solved so far but there are some hopes that it may be solved in non-perturbative approach.
We do not go to the discussion of this problem because, as we explained in the introduction, we will consider the effective gravity theory which is not renormalizable due to presence of the dark energy $F(R)$-term added by hands, so this is just classical modified gravity. Furthermore, effectively Friedmann equations of our complete theory are just the same as for corresponding ghost-free $F(R)$-gravity obtained from our theory by dropping the Weyl-squared term which does not give any contribution to the equations of motion.

In this paper we will use the approach of Ref.~\cite{B} (see also the references therein), where, by starting from quadratic higher-derivative gravity, it is possible to obtain a renormalizable model by computing the one-loop divergences of the theory.

Quadratic higher-derivative gravity models are quite interesting, since they represent a modification of Einstein's gravity at high curvature, where the phenomenology of early-time inflation can be strictly connected with them. The very general action of this theory is given by,
\begin{equation}
I=\int_\mathcal{M} d^4 x\sqrt{-g}\left(\frac{R}{\kappa_0^2}-\tilde\Lambda+a R^2
+b R_{\mu\nu}R^{\mu\nu}+c R_{\mu\nu\xi\sigma}R^{\mu\nu\xi\sigma}+d\Box
R\right)\,,\label{azione0}
\end{equation}
where $g$ is the determinant of the metric tensor $g_{\mu\nu}$, $\mathcal{M}$
is the space-time manifold, and $\Box\equiv g^{\mu\nu}\nabla_{\mu}\nabla_{\nu}$ the
covariant d'Alembertian, ${\nabla}_{\mu}$ being the covariant derivative
operator associated with the metric $g_{\mu \nu}$.
The Hilbert-Einstein action is given by the Ricci scalar $R$, while  $R^2\,,R_{\mu\nu}R^{\mu\nu}\,,R_{\mu\nu\xi\sigma} R^{\mu\nu\xi\sigma}$ and $\Box R$ are
the higher curvature corrections to GR,  $R_{\mu\nu}$ and $R_{\mu\nu\xi\sigma}$ being the
Ricci  and the Riemann tensors, respectively. Here, $0<\kappa_0^2$ encodes the mass scale of the theory and $a, b, c, d$ are constant parameters. Finally, $\tilde\Lambda$ is a cosmological constant term, which should not be confused with the cosmological constant $\Lambda$ for dark energy.
If we here introduce the Gauss-Bonnet four-dimensional topological invariant, $G$, and the square of the Weyl tensor, $C^2$\,,
\begin{equation}
G=R^2-4R_{\mu\nu}R^{\mu\nu}+R_{\mu\nu\xi\sigma}R^{\mu\nu\sigma\xi}\,,\quad
C^2=\frac{1}{3}R^2-2R_{\mu\nu}R^{\mu\nu}+R_{\xi\sigma\mu\nu}R^{\xi\sigma\mu\nu}\,,\label{Gauss}
\end{equation}
we can write
\begin{equation}
R_{\mu\nu}R^{\mu\nu}=\frac{C^2}{2}-\frac{G}{2}+\frac{R^2}{3}\,,\quad
R_{\mu\nu\xi\sigma}R^{\mu\nu\xi\sigma}=2C^2-G+\frac{R^2}{3}\,.
\end{equation}
As the Gauss-Bonnet and the surface term $\Box R$ do not contribute to the dynamical field equations of the model, we can drop them down from the action, which will result in terms of $R/\kappa_0^2$, $\tilde\Lambda$, $R^2$ and $C^2$ only.

Using the results of the one-loop calculatios in the above theory, we can now proceed with it RG improvement, in analogy with RG-improved calculations carried out in quantum field theory in curved space-time~\cite{El, rginfl}. In this way (see, also, Ref.~\cite{qg1}), we obtain a RG-improved action for higher-derivative quantum gravity, which reads
\begin{equation}
I=\int_\mathcal{M}d^4\sqrt{-g}\left[\frac{R}{\kappa^2(t')}-\tilde\Lambda(t')-\frac{\omega(t')}{3\lambda(t')}R^2+
\frac{1}{\lambda(t')}C^2+ f_{\text{DE}}(R)+\mathcal L_m\right]\,.\label{action0}
\end{equation}
Note that here, to the RG-improved action we have added $\mathcal L_m$, which is the Lagrangian corresponding to standard matter, and the function of the Ricci scalar $f_\text{DE}(R)$, which has been introduced by hand to support the late-time cosmic acceleration. The precise role of this term will be discussed in Section {\bf V}. In this way we unify in a single model the higher curvature corrections, which account for quantum gravity effects for inflation, with a phenomenological $F(R)$-gravity for the dark energy era. Notice also that QG corrections in $\mathcal L_m$ can be safely neglected, since inflation is assumed to occur owing to the presence of purely gravitational terms.

The running coupling constants $\kappa^2\equiv\kappa^2(t')$, $\tilde\Lambda\equiv\tilde\Lambda(t')$, $\lambda\equiv\lambda(t')$ and $\omega\equiv\omega(t')$ are obtained from one-loop quantum gravity corrections and, thus,  we deal with a renormalization-group-improved effective action from multiplicatively-renormalizable
quantum gravity. This kind of Lagrangian has been investigated in several papers~\cite{El,rginfl}. In its simplest formulation
~\cite{El}, the RG-improved effective action follows from
the solution of the RG equation applied to the complete effective action of the
multiplicatively renormalizable theory. As a final result, the one-loop coupling
constants are expressed in terms of the log term of a
characteristic mass scale in the theory, namely
\begin{equation}
t'=\frac{t'_0}{2}\log\left[\frac{R}{R_0}\right]^2\,,\quad 0<t_0'\,,
\label{t'}
\end{equation}
where $t_0'$ is a positive number and $R_0$ is the curvature at which the quantum gravity effects disappear.
The running coupling constants for the gravitational action (\ref{action0}) obey  the one-loop RG
equations~\cite{B,Fradkin},
\begin{eqnarray}
\frac{d\lambda}{d
t'}&=&-\beta_2\lambda^2\,,\quad
\frac{d\omega}{d
t'}=-\lambda(\omega\beta_2+\beta_3)\,,\quad
\frac{d\kappa^2}{d
t'}=\kappa^2\gamma\label{kappaeq}\,,
\quad
\frac{d\tilde\Lambda}{d
t'}=\frac{\beta_4}{\left(\kappa^2\right)^2}-2\gamma\tilde\Lambda\,,
\label{runeq}
\end{eqnarray}
and are closely connected with the $\beta$-functions $\beta_{2,3,4}$ and $\gamma$, namely
\begin{equation}
\beta_2=\frac{133}{10}\,,\quad\beta_3=\frac{10}{3}\omega^2+5\omega+\frac{5}{12}\,,\quad
\beta_4=\frac{\lambda^2}{2}\left(5+\frac{1}{4\omega^2}\right)+\frac{\lambda}{3}\kappa^4\tilde\Lambda\left(20\omega+15-\frac{1}{2\omega}\right)\,,
\quad
\gamma=\lambda\left(\frac{10}{3}\omega-\frac{13}{6}-\frac{1}{4\omega}\right)\,.\label{betabeta}
\end{equation}
We observe that the renormalization procedure does not allow us to set $\tilde\Lambda=0$.
We also should note that, when the coefficients in the action (\ref{azione0}) are not constant, the Gauss-Bonnet and the $\Box R$-term contribute to the dynamics of the model, resulting into additional RG equations. As stated before, in this work we will
use the simplest renormalized theory of Ref.~\cite{B} with $R^2$ and Weyl corrections only, but in the next chapter we will offer a detailed comparison with the extended model in FRW space-time.

The important remark is in order. Despite to the fact that standard scalar
radiation/matter fields quickly are shifted away during the early-time
accelerated expansion, their presence may slightly affect the behavior of the
running coupling constants in (\ref{runeq})--(\ref{betabeta}). In Ref.~\cite{qg2} the scalar electrodynamics model non-minimally coupled with higher-derivative  gravity has been analyzed in the framework
of quadratic gravity by taking into account the whole form of quantum
corrections including the matter. The coupling with gravity leads to an effective field
potential in the Lagrangian together with additional terms in the beta
functions (\ref{betabeta}). However, if one considers simple models where the matter
fields are decoupled from gravity, the corrections to the RG equations
above can be ignored as curvature terms are dominant if compare with scalar potential similar to Starobinsky inflation with matter.
The matter Lagrangian may contain some log-terms of the field whose
contribution appears in the perturbed equations, but leads to negligible
corrections in the spectral index and in the tensor-to-scalar ratio of the
model. For these reasons, assuming Standard Model for the matter
Lagrangian, we can use the results in (\ref{runeq})--(\ref{betabeta}). We explicitly checked that account of matter contribution (scalar theory and/or scalar electrodynamics) in RG improved effective action (\ref{action0}) does not lead to any qualitative changes in the results of this paper.

From the first equation in (\ref{runeq}), we immediately get:
\begin{equation}
\lambda(t')=\frac{\lambda(0)}{1+\lambda(0)\beta_2 t'}\,,\quad 0<\lambda(0)\,,\label{lambda}
\end{equation}
where $\lambda(0)$ is a positive integration constant, corresponding to the value of
$\lambda(t')$ at $t'=0$, namely $R=R_0$ in (\ref{t'}).

The equation for $\omega$ exhibits two fixed points, namely $\omega_1\simeq -0.02$ and $\omega_2\simeq -5.47$, but only the first of them is stable, being an attractor of the system for large values of $t'$, namely when $R_0\ll R$~\cite{Fradkin}. We can easily demonstrate this extreme by considering the behavior of
$\omega(t')=\omega_{1,2}+\delta\omega(t')$, $|\delta\omega(t')/\omega_{1,2}|\ll 1$,
around the fixed points, namely
\begin{eqnarray}
\hspace{-1cm}\frac{d\omega(t')}{d
t'}&\simeq&-\lambda(t')\left(\frac{20}{3}\omega+\frac{183}{10}\right)\vert_{\omega_{1,2}}\delta\omega(t')
-\lambda(t')^2\beta_2\left(\frac{d
t'}{d\omega(t')}\right)\left(\frac{10}{3}\omega^2+\frac{183}{10}\omega+\frac{5}{12}\right)\vert_{\omega_{1,2}}\delta\omega(t')\nonumber\\
&=&-\lambda(t')\left(\frac{20}{3}\omega+\frac{158}{5}\right)\vert_{\omega_{1,2}}\delta\omega(t')\,,\label{omegapert}
\end{eqnarray}
which leads to
\begin{equation}
\omega(t')=\omega_{1,2}+\frac{c_0}{(1+\lambda(0)\beta_2 t')^{q}}\,,\quad
q=\frac{1}{\beta_2}
\left(\frac{20}{3}\omega+\frac{158}{5}\right)\vert_{\omega_{1,2}}\,,\label{inter}
\end{equation}
where $|c_0|\ll 1$ is a constant and we have used
(\ref{lambda}). Thus, the solution is stable at $t'\rightarrow\infty$ for $\omega_1$ with $q\simeq2.37$, while for $\omega_2$ with $q\simeq-0.37$ it diverges. This means that, for large values of $t'$, the function $\omega(t')$ tends to the attractor at $\omega=\omega_1$. By taking into account that $0<d\omega(t')/d t'$, when $0<\lambda(0)$ and $\omega_2<\omega<\omega_1$, the function
$\omega(t')$ grows up with $t'$ and approaches  $\omega_1$, as
\begin{equation}
\omega(t')=\omega_{1}+\frac{c_0}{(1+\lambda(0)\beta_2 t')^{p}}\,,\quad
p=\left(\frac{10}{3}\right)\frac{(\omega_1-\omega_2)}{\beta_2}\simeq
1.36\,,\label{omega1}
\end{equation}
where we have taken the average value of $\omega$ between $\omega_1$ and $\omega_2$.
In this paper we will set $c_0=0$, so that
\begin{equation}
\omega(t')=\omega_1=-0.02\,.
\label{omega0.02}
\end{equation}
Note that the term $R^2$ gives a positive contribution inside the gravitational action and helps to avoid finite-time  future singularities at large curvature.

Now, it is possible to get the form of $\kappa^2$ from the third equation in (\ref{runeq}),
\begin{equation}
\kappa^2(t')=\kappa_0^2(1+\lambda(0)\beta_2 t')^{Z/\beta_2}\,,\quad
Z=\left(\frac{10}{3}\omega_1-\frac{13}{6}-\frac{1}{4\omega_1}\right)=
10.27\,,
\end{equation}
where $\kappa_0^2=\kappa^2(0)$ corresponds to the mass scale of the theory at small curvature, namely the Planck mass $M_{\text{Pl}}^2$,
\begin{equation}
\kappa_0^2=\frac{16\pi}{M_{\text{Pl}}^2}\,,\quad M_\text{Pl}^2=1.2\times 10^28 \text{eV}\,.\label{kappa0exp}
\end{equation}
Finally, we derive the form of $\tilde\Lambda$ from the last equation in (\ref{runeq}), which reads
\begin{equation}
\frac{d(\kappa^4\tilde\Lambda)}{dt'}=\beta_4\equiv
\frac{\lambda(t')^2}{2}\left(5+\frac{1}{4\omega(t')^2}\right)+\lambda(t')\left(\kappa^4(t')\tilde\Lambda(t')\right)\left(\frac{20}{3}\omega(t')+5-\frac{1}{6\omega(t')}\right)\,.
\end{equation}
When $\omega=\omega_1$ is constant, this equation can be solved with respect to the dimensionless function $\kappa^4\tilde\Lambda$, and leads to
\begin{equation}
\kappa^4\tilde\Lambda=-\frac{3\lambda(0)(1+20\omega_1^2)}{4\omega_1(1+\lambda(0)\beta_2
t')(-1+30\omega_1+6\beta_2\omega_1+40\omega_1^2)}
+\kappa_0^4\tilde\Lambda_0(1+\lambda(0)\beta_2 t')^{W/\beta_2}\,,\quad
W=\frac{20}{3}\omega_1+5-\frac{1}{6\omega_1}=13.2\,,
\label{fullLambda}
\end{equation}
where we have used (\ref{lambda}). Since we would like to completely avoid the quantum induced effects at small curvature, we require that $\tilde\Lambda= 0$ when $t'= 0$, by fixing the integration constant $\tilde\Lambda_0$ as
\begin{equation}
\tilde\Lambda_0=\frac{3\lambda(0)}{\kappa_0^4}\left(
\frac{(1+20\omega_1^2)}{4\omega_1(-1+30\omega_1+6\beta_2\omega_1+40\omega_1^2)}
\right)\simeq -\frac{11\times\lambda(0)}{\kappa_0^4}\,.
\label{parLambda0}
\end{equation}
On the other hand, due to the fact that $0<W$, at large curvature 
\begin{equation}
\tilde\Lambda\simeq
\frac{\tilde\Lambda_0}{(1+\lambda(0)\beta_2
t')^{X/\beta_2}}\simeq \frac{\tilde\Lambda_0}{(1+\lambda(0)\beta_2
t')^{0.55}}
\,,\quad X=(2Z-W)\simeq 7.34\,,\label{parLambda}
\end{equation}
and the cosmological constant from the RG improved effective action tends to decrease.

A general remark is here in order. Quantum corrections must disappear towards the end of inflation, when $R\leq R_0$. For this reason the boundary term $\lambda(0)$ must be chosen according to the condition
\begin{equation}
\lambda(0)\ll |\frac{1}{\beta_2t_0'\log[4\Lambda/R_0]}|\,,
\label{condlambda0}
\end{equation}
where $R=4\Lambda$ corresponds to the de Sitter (final) curvature attractor of the dark energy epoch of our Universe today, $\Lambda$ being the cosmological constant. Only in this way can we be sure that quantum corrections are negligible in the whole curvature range $4\Lambda<R<R_0$. Thus, the parametrization in (\ref{parLambda}) guarantees that the cosmological constant term $\Lambda(t')$ does not play any significative role, neither at high curvature nor at small curvature, and thus we can neglect its contribution.

In the next chapter, the possibility to get an early-time inflation from our gravitational model will be discussed. Contributions from the modified function $f_{\text{DE}}(R)$ and the matter Lagrangian inside the action turn out to be negligible, and we will consider inflation as a manifestation of high-curvature corrections to GR, which take into account quantum effects.


\section{Early-time inflation in higher-derivative gravity}

Let us consider the general form of flat FRW space-time
\begin{equation}
ds^2=-N(t)^2dt^2+a(t)^2(dx^2+dy^2+dz^2)\,,\label{metric}
\end{equation}
where $a\equiv a(t)$ is the scale factor of the Universe and $N(t)$ is an {\it einbein} function of the cosmological time.
If we take the variation of the Weyl term, as
\begin{equation}
\delta I_{C^2}=\frac{1}{\lambda(t')}\delta\left(\sqrt{-g}C^2\right)+
\left(\sqrt{-g}C^2\right)\delta\left(\frac{1}{\lambda(t')}\right)\,,
\end{equation}
we immediately note that on the FRW metric (\ref{metric}),
\begin{equation}
C^2=0\,,\quad \frac{1}{\lambda(t')}\delta\left(\sqrt{-g}C^2\right)=0\,,
\end{equation}
so that the square of the Weyl tensor does not enter into the
Friedmann field equations of the theory. In what follows, we will fix the usual gauge as $N(t)=1$.

The evolution of the model in the vacuum at high curvature is governed by the first Friedmann-like equation. If we neglect the contribution of $f_\text{DE}(R)$ and $\mathcal L_m$ in (\ref{action0}), the first Friedmann-like equation reads~\cite{qg1}
\begin{eqnarray}
0&=&
\frac{6H^2}{\kappa^2(t')}-\frac{6H}{(\kappa^2(t'))^2}\frac{d\kappa^2(t')}{dt'}
\left(\frac{t_0'\dot R}{R}\right)
+\frac{\omega(t')}{3\lambda(t')}
\left[6R\dot H
-12H\dot R\right]
-12 H\frac{d}{d
t'}\left(\frac{\omega(t')}{3\lambda(t')}\right)\dot
R t_0'\nonumber\\&&
+6\left(H^2+\dot H\right)\Delta(t')\frac{t_0'}{R}-6H
\left[\frac{d\Delta(t')}{d t'}\left(\frac{t'_0}{
R}\right)^2-\Delta(t')\frac{t_0'}{R^2}\right]\dot R-\tilde\Lambda(t')\,,
\label{NN}
\end{eqnarray}
with
\begin{equation}
\Delta(t')=\left[\frac{R}{(\kappa^2(t'))^2}\frac{d\kappa^2(t')}{d
t'}+R^2\frac{d}{d
t'}\left(\frac{\omega(t')}{3\lambda(t')}\right)+\frac{d\tilde\Lambda(t')}{d
t'}\right]\,.
\label{Delta}
\end{equation}
In the above expressions, the dot denotes  time derivative, and the Ricci scalar is given by,
\begin{equation}
R=12H^2+6\dot H\,,\quad H=\frac{\dot a}{a}\,,\label{Ricci}
\end{equation}
with $H$ the Hubble parameter.

Inflation is described by a (quasi) de Sitter solution, where the Hubble parameter is almost a constant, namely $H=H_\text{dS}$.
On the de Sitter solution $R_\text{dS}=12 H_\text{dS}^2$, the system leads to
\begin{eqnarray}
\frac{6H_\text{dS}^2}{\kappa^2(t')}
+\Delta(t')\frac{t_0'}{2}-\tilde\Lambda(t')=0\,,\quad
\Delta(t')=\left[\frac{12H_\text{dS}^2}{(\kappa^2(t'))^2}\frac{d\kappa^2(t')}{d
t'}+144H_\text{dS}^4\frac{d}{d
t'}\left(\frac{\omega(t')}{3\lambda(t')}\right)+\frac{d\tilde\Lambda(t')}{d
t'}\right]\,.
\label{dSeq}
\end{eqnarray}
It is interesting to compare these expressions with the ones from higher-derivative quantum gravity including the Gauss-Bonnet $G$ and  $\Box R$-terms. As we observed in the preceding section, when one deals with running coupling constants, the most general form of a quadratic higher-derivative theory is given by (\ref{action0}) plus the following gravitational part
\begin{equation}
I_{G\,,\Box R}=-\int_\mathcal M d^4 x\sqrt{-g} \left[\gamma(t') G-\zeta(t')\Box
R\right]\,,\label{new}
\end{equation}
with $\gamma\equiv\gamma(t')$ and $\zeta\equiv\zeta(t')$ functions of the scale parameter $t'$. In this case, the first Friedmann-like equation of the model reads~\cite{qg1}
\begin{eqnarray}
0&=&
\frac{6 H^2}{\kappa^2(t')}-\frac{6
H}{(\kappa^2(t'))^2}\frac{d\kappa^2(t')}{d t'}\left(\frac{t_0' \dot
R}{R}\right)
+\frac{\omega(t')}{3\lambda(t')}
\left[6R\dot H
-12H\dot R\right]
-12 H\frac{d}{d t'}\left(\frac{\omega(t')}{3\lambda(t')}\right)\left(\dot R
t_0'\right)\nonumber\\&&
+6\left(H^2+\dot H\right)\Delta(t')\frac{t_0'}{R}-6 H
\left[\frac{d\Delta(t')}{d
t'}\left(\frac{t_0'}{R}\right)^2-\Delta(t')\frac{t_0'}{R^2}\right]\dot R
-24 H^3\frac{d\gamma(t')}{d t'}\frac{t'_0\dot R}{R}-6
H\left[\frac{d\gamma(t')}{d t'}\frac{t_0'\dot G}{R}\right]\nonumber\\&&
-3\mathcal A\dot R^2
-2\mathcal B\dot R^2 R
+6\frac{d}{d t}\left[2\mathcal A\left(4H^2+3\dot H\right)\dot R+\mathcal B
H\dot R^2\right]
+18 H\left[2\mathcal A\left(4H^2+3\dot H\right)\dot R+\mathcal B H\dot
R^2\right]\nonumber\\
&&-36\left(3H^2+\dot H\right)\mathcal A\,H\dot R-72H\frac{d}{d t}\left(\mathcal
A H \dot R\right)-12\frac{d^2}{d t^2}
\left(\mathcal A H \dot R\right)-\tilde\Lambda(t')
\,,
\label{NN2new}
\end{eqnarray}
where $\Delta(t')$ is still given by (\ref{Delta}), the Gauss-Bonnet on FRW space-time corresponds to
\begin{equation}
G=24H^2\left(H^2+\dot H\right)\,,
\end{equation}
and
\begin{equation}
\mathcal A=\left(\frac{d\zeta(t')}{d t'}\frac{t'_0}{R}\right)\,,\quad
\mathcal B=\left[\frac{d^2\zeta(t')}{d
t'^2}\left(\frac{t'_0}{R}\right)^2-\frac{d\zeta(t')}{d
t'}\frac{t'_0}{R^2}\right]\,.
\end{equation}
We see that, when $\gamma\,,\zeta$ are constant, we recover (\ref{NN}).
For the de Sitter solution $H=H_\text{dS}$ we have now
\begin{eqnarray}
\frac{6H_\text{dS}^2}{\kappa^2(t')}
+\Delta(t')\frac{t_0'}{2}-\tilde\Lambda(t')+12H_\text{dS}^4\frac{d\gamma}{d t'}t_0'=0\,,
\label{dSeq2}
\end{eqnarray}
and (of course) only the Gauss-Bonnet gives an additional contribution with respect to the case in (\ref{dSeq}).

Generally speaking, for large values of the Hubble parameter, when we neglect $\tilde\Lambda$, Eq.~(\ref{dSeq}) assumes the form
\begin{equation}
-\alpha(t') H_\text{dS}^4+\beta(t') H_\text{dS}^2\simeq 0\,,\label{gen0}
\end{equation}
where $\alpha$ and $\beta$ are dimensionfull (positive) functions of the scale parameter $t'$. Thus, the de Sitter solution reads,
\begin{equation}
H_\text{dS}^2\simeq\frac{\beta(t')}{\alpha(t')}\,,\quad t'=\frac{t_0'}{2}\log\left[\frac{R_\text{dS}}{R_0}\right]^2\,.
\end{equation}
If we add to the model the Gauss-Bonnet contribution, the first Friedmann-like equation for constant Hubble parameter (\ref{dSeq2}) takes the (asymptotic) form:
\begin{equation}
\left(-\alpha(t') +12\frac{d\gamma}{d t'}t_0'\right)H_\text{dS}^4+\beta(t') H_\text{dS}^2\simeq 0\,,
\end{equation}
leading to the solution,
\begin{equation}
H_\text{dS}^2\simeq\frac{\beta(t')}{\left[\alpha(t')-12(d\gamma/dt')t_0'\right]}\,.
\end{equation}
We conclude that a contribution of the Gauss-Bonnet kind in the gravitational action as $\sim-\gamma(t')G$ increases the curvature of inflation if $0<d\gamma/dt'$, and vice-versa, it decreases the curvature of the inflationary Universe provided that $d\gamma/dt'<0$.\\
\\
Let us return to the simplified action (\ref{action0}) with (\ref{NN})--(\ref{dSeq}). By using the set of equations (\ref{runeq})--(\ref{betabeta}), one derives, from (\ref{dSeq}),
\begin{eqnarray}
0&=&
\frac{6 H^2}{\kappa ^2}-\frac{t_0'}{48 (\kappa ^2)^2 \omega ^2} \left(480 H^4
(\kappa^2)^2 \omega ^2 (4 \omega  (2
      \omega +3)+1)+24 \kappa ^2 \lambda  \omega   H^2 \left(-40 \omega
^2+26 \omega
      +3\right)
\right.
\nonumber\\&&
\left.
-3 \lambda ^2 \left(20 \omega
      ^2+1\right)\right)\,,
\label{dSeq22}
\end{eqnarray}
where the functions $\lambda\,,\omega\,,\kappa^2$ are assumed to
be constant with respect to time.

By taking into account the expressions of $\kappa^2\,,\lambda\,,\omega$ and $\tilde\Lambda$ derived in the preceding section (note that $\alpha(t')$ in (\ref{gen0}) turns out to be constant), by taking $\kappa_0^4\tilde\Lambda\ll 1$ (see (\ref{parLambda0})--(\ref{parLambda}) with $\lambda_0\ll 1$),
we derive, in the limit
$1\ll t'$ when the quantum corrections are relevant,
\begin{equation}
H_\text{dS}^2\kappa_0^2\simeq
\frac{0.107}{t_0'(\lambda(0)t')^{0.77}}\,.\label{HdS}
\end{equation}
In fact, our de Sitter solution emerges from the one-loop corrections encoded in (\ref{Delta}). In  Starobinsky's inflationary scenario, where the coefficients of the action are constant and come from the trace-anomaly~\cite{Staro},  the $R^2$-term alone supports the de Sitter solution of inflation, while the Hilbert-Einstein term $R/\kappa^2$ permits to slowly exit from the accelerated phase (this role can be played also by different power functions $R^\delta$ with $\delta<2$~\cite{mioStaro}). However even in the Starobinsky model, when one takes the asymptotic limit $R/\kappa^2\ll R^2$ in the Friedmann equation ($0\simeq R_\text{dS}^2-12R_\text{dS} H_\text{dS}^2$), the mass scale of the theory is implicitly considered to be smaller than the Planck mass, in order to avoid super-Planckian curvatures, implying a sort of ``running'' mechanism (for some recent works on $R^2$-gravity in this respect, see Refs.~\cite{Rinaldi}).
\\
\\
We will now proceed with the investigation of the graceful exit from inflation.
Let us consider a small perturbation around the quasi de Sitter solution described by (\ref{HdS}), namely
\begin{equation}
H=H_\text{dS}+\delta H(t)\,,\quad |\delta H(t)/H_\text{dS}|\ll1\,.\label{approxH}
\end{equation}
In the limit $1\ll t'$, if we neglect the contribution of $\tilde\Lambda$, Eq.~(\ref{NN})  reads
\begin{eqnarray}
&&\hspace{-1cm}0=(\kappa_0\dot\delta H)\left[t_0'\left(
\left(H_{\text{dS}}\kappa_0\right)^2
\left(34.344 -\frac{0.913
      t_0'}{t'}\right)+\frac{0.001 t'+0.003 t_0'}{t'^3
\left(H_{\text{dS}}\kappa_0\right)^2 (\lambda(0)
      t')^{1.54}}+\frac{0.346 t_0'-0.086 t'}{t'^2 (\lambda(0)
t')^{0.77}}\right)
+19.152 t'
      \left(H_{\text{dS}}\kappa_0\right)^2\right]
\nonumber\\&&
\hspace{1cm}+\frac{(\kappa_0^2 \ddot\delta H )}{t'^3
\left(H_{\text{dS}}\kappa_0\right)^3
}\left[\phantom{\frac{0}{0}}t'^2
\left(H_{\text{dS}}\kappa_0\right)^4
      \left(6.384 t'^2+t_0' (11.448 t'-0.228 t_0')\right)
\right.
\nonumber\\
&&\left.
\hspace{1cm}
      -\frac{0.043 t'^2 t_0' \left(H_{\text{dS}}\kappa_0\right)^2}{(\lambda(0)
t')^{0.77}}+ \frac{0.001 t_0'^2}{(\lambda(0) t')^{1.54}}+\frac{0.087 t'
      t_0'^2 \left(H_{\text{dS}}\kappa_0\right)^2 }{(\lambda(0) t')^{0.77}}
+\frac{2\times 10^{-4} t' t_0' }{(\lambda(0)
      t')^{1.54}}\phantom{\frac{0}{0}}\right]\nonumber\\&&
\hspace{1cm}+\left(H_{\text{dS}}\kappa_0\right)
\delta H
      \left[\frac{0.223}{(\lambda(0) t')^{0.77}}+\frac{0.172 \lambda(0)
      t_0'}{(\lambda(0) t')^{1.77}}-30.528 t_0'
\left(H_{\text{dS}}\kappa_0\right)^2\right]\,,\label{sistpert}
\end{eqnarray}
and, for $1\ll t'$, it leads to~\cite{qg1}
\begin{equation}
D_0\delta H(t)+
t'[19.152 (H_\text{dS}\kappa_0)(\kappa_0\dot \delta
H(t))+6.384(\kappa_0^2\ddot\delta H(t))]\simeq 0\,,\label{stabeq}
\end{equation}
with
\begin{equation}
D_0=
      \left[\frac{0.223}{(\lambda(0) t')^{0.77}}-30.528 t_0'
\left(H_{\text{dS}}^2\kappa_0^2\right)\right]\simeq-28.444
t_0'
\left(H_{\text{dS}}^2\kappa_0^2\right)
\,.\label{D0}
\end{equation}
The solution of Eq.~(\ref{stabeq}) is 
\begin{equation}
\delta H(t)= h_\pm\exp\left[A_\pm t\right]\,,
\quad
A_\pm=
\left[\frac{H_{\text{dS}}}{2}\left(-3\pm\sqrt{9-\frac{0.627
D_0}{(H_\text{dS}^2\kappa_0^2)t'}}\right)\right]\,,
\quad
|h_{\pm}/H_\text{dS}|\ll 1\,,\label{solpert}
\end{equation}
where $h_\pm$ are  integration constants corresponding to the plus and minus
signs inside $A_\pm$, respectively. Since $D_0$ is negative, the solution turns out to to be unstable, namely, for $1\ll t'$,
\begin{equation}
\delta H(t)\simeq h_+\text{e}^{A_+ t}\,,\quad A_+\simeq 1.486\left(\frac{H_\text{dS}t_0'}{t'}\right)\,,\label{deltaH}
\end{equation}
with $h_+<0$ to make the Hubble parameter decreasing.

If we introduce $t_\text{i}$ and $t_\text{e}$ as the initial and the final time of the early-time acceleration, respectively, we can set $h_+=-H_\text{dS}\text{e}^{-A_+t_\text{e}}$, and thus obtain
\begin{equation}
H=H_\text{dS}\left(1-\text{e}^{A_+ (t-t_\text{e})}\right)\,,\quad t_\text{i}\ll t_\text{e}\,,
\end{equation}
so that the Hubble parameter tends to vanish at the end of inflation.

The number of $e$-folds is a valid parametrization frequently used in the study of early-time inflation. It is defined as
\begin{equation}
N=\ln \left(\frac{a(t_\mathrm{e})}{a (t)}\right)\,.
\label{Nfold}
\end{equation}
By taking into account that $a(t)\sim \text{e}^{H_\text{dS}t}$, we get
\begin{equation}
t=t_\text{e}-\frac{N}{H_\text{dS}}\,.
\label{tN}
\end{equation}
Thus, the Hubble parameter during inflation behaves as
\begin{equation}
H\simeq H_\text{dS}\left(1-\text{e}^{-\frac{A_+ N}{H_\text{dS}}}\right)\,.
\label{HN}
\end{equation}
When $0\ll N$  the Hubble parameter is given by the de Sitter solution $H\simeq H_\text{dS}$, while, when $N\rightarrow 0$, the Hubble parameter decreases and goes to zero (i.e. $R_0\ll R_\text{dS}$), allowing the model to exit from the inflationary phase.\\
\\
The total $e$-fold number of the corresponding inflation
\begin{equation}
\mathcal N=\ln \left(\frac{a (t_\mathrm{e})}{a(t_\mathrm{i})}\right)\simeq H_\text{dS}(t_\text{e}-t_\text{i})\,,
\label{N}
\end{equation}
must be large enough in order to appropriately lead to the thermalization
of our observable Universe and to solve the problem of the initial conditions of the Friedmann Universe, too. In general it is required that $60\leq\mathcal N$. Moreover, given that when $N=\mathcal N$ it must be $H\simeq H_\text{dS}$, we can identify $H_\text{dS}/A_+$ as the minimal value of $e$-folds $\mathcal N_\text{min}$ at which inflation can start, namely
\begin{equation}
60\leq\mathcal N_\text{min}=\frac{H_\text{dS}}{A_+}\leq \mathcal N\,.
\label{320}
\end{equation}
It turns out that (\ref{HN}) can be rewritten as
\begin{equation}
H\simeq H_\text{dS}\left(1-\text{e}^{-\frac{N}{\mathcal N_\text{min}}}\right)\,.
\label{HN2}
\end{equation}
We observe that $\mathcal N_\text{min}$ encodes the curvature expansion rate of inflation as
\begin{equation}
\mathcal N_\text{min}\simeq (1.486)^{-1}\log\left[\frac{R_\text{dS}}{R_0}\right]\,,
\label{330}
\end{equation}
where we have used (\ref{deltaH}) with (\ref{t'}).
Once the curvature expansion rate $R_\text{dS}/R_0$ is fixed,  by taking into account Eq.~(\ref{HdS}), we obtain
a relation between the curvature at the time of inflation and the boundary parameter $\lambda(0)$,
\begin{equation}
R_\text{dS}=\frac{1.284}{(t_0'\kappa_0^2)\left(\lambda(0)t_0'\log[R_\text{dS}/R_0]\right)^{0.77}}\,.\label{RdSlambda}
\end{equation}
In the next section we will investigate the cosmological perturbations left at the end of inflation, and we will derive the spectral index and the tensor-to-scalar ratio of the model. Accurate comparison with astronomical data will establish the value of the curvature expansion rate $R_\text{dS}/R_0$.

\section{Cosmological perturbations during inflation}

During the inflationary stage the Hubble parameter slowly decreases in the so called ``slow-roll approximation'' regime, provided the following conditions are met
\begin{equation}
|\frac{\dot H}{H^2}|\ll 1\,,\quad |\frac{\ddot H}{H\dot H}|\ll
1\,.\label{slowrollcondition}
\end{equation}
In particular,  the slow-roll parameter  $\epsilon$,
\begin{equation}
\epsilon=-\frac{\dot H}{H^2}\equiv \frac{1}{H}\frac{d H}{d N}\,,
\end{equation}
must be small and positive during inflation, while it tends towards one when the early-time acceleration ends.
Using (\ref{HN2}), we obtain
\begin{equation}
\epsilon\simeq\frac{\text{e}^{-\frac{N}{\mathcal N_\text{min}}}}{\mathcal N_\text{min}}\,.
\label{epsilonN}
\end{equation}
One of the most important prediction of inflation consists in the description of the anisotropies of our Universe at galactic scale. In this respect, perturbation theory  is the key mechanism to calculate the inhomogeneities left at the end of the primordial accelerated expansion, and leads to
the derivation of the spectral index and of the tensor-to-scalar ratio for scalar and tensorial perturbations, respectively. Therefore, only if these indexes fit the inferred values in our observable Universe will the theory be considered to be viable and to lead to a realistic description of inflation.

An important remark is in order. If one considers the full form of the renormalizable action (\ref{action0}), the Weyl term appears. As we recalled before, it actually does not contribute to the Friedmann-like equations of the theory: neither inflation itself, nor the graceful exit phase,
$e$-fold number (\ref{N}), nor the $\epsilon$ slow-roll parameter (\ref{epsilonN}) depend on it.
However, when one introduces the cosmological perturbations around the FRW metric, the Weyl term plays a role in the perturbed equations~\cite{Derue, corea1, corea2}. In Refs.~\cite{corea1} it has been shown that in pure Weyl conformal gravity the scalar perturbations do not propagate in the de Sitter background, while the vector and the tensor power spectra are constant, as a consequence of the invariance of the Weyl tensor with respect to conformal transformations. This behavior seems to be confirmed in Weyl invariant scalar tensor theories~\cite{corea2}. In our case, however, the Weyl term may change the evolution of the cosmological perturbations of the model. For example, if one starts with scalar perturbations in the Newton's gauge,
\begin{equation}
ds^2=-\left(1+2\Phi(t,{\bf x})\right)dt^2+a(t)^2\left(1-2\Psi(t,{\bf x})\right)(dx^2+dy^2+dz^2)\,,
\end{equation}
with $\Phi\equiv \Phi(t,{\bf x})$ and $\Psi\equiv\Psi(t,{\bf x})$ scalar functions of the space-time coordinates, we derive $\sqrt{-g}C^2=(4/3)\left[\nabla^2(\Phi+\Psi)\right]^2$. In the background of Einstein's gravity one has $\Phi=\Psi$, while in a $F(R)$-theory of gravity $\Phi=-\dot\Psi/(H+\dot F_R(R)/(2F_R(R)))$, and we see that the square of the Weyl tensor is here different from zero.

It lies beyond the scope of our work to investigate these implications of the Weyl tensor in the perturbative theory of $F(R)$-gravity, namely because in our action the square of the Weyl tensor $C^2$ is coupled with the Ricci scalar at the inflationary scale, rendering the system very involved. Cosmological perturbation theory in the presence of the Weyl term is still a debated subject (see e.g. Ref.~\cite{Man}).
In what follows, we will neglect its contribution in  dealing with a $F(R)$-gravity model.
Thus, the spectral index $n_s$ and the tensor-to-scalar ratio $r$ read~\cite{corea},
\begin{equation}
(1-n_s)\simeq -\frac{2}{\epsilon}\frac{d\epsilon}{d N}\,,\quad
r\simeq 48\epsilon^2\,,\label{spectralMG}
\end{equation}
where for the tensor-to-scalar ratio we have used second order corrections (the first order ones simply vanish). By using (\ref{epsilonN}), we immediately get
\begin{equation}
(1-n_s)=\frac{2}{\mathcal N_\text{min}}
\,,
\quad
r=\frac{48}{\mathcal N_\text{min}^2}
\text{e}^{-\frac{2 \mathcal N}{\mathcal N_\text{min}}}\,,
\end{equation}
where we set $N=\mathcal N$ during inflation.
Recent analysis of Planck data~\cite{Planck} constraint these quantities as
$n_{\mathrm{s}} = 0.968 \pm 0.006\, (68\%\,\mathrm{CL})$ and
$r < 0.11\, (95\%\,\mathrm{CL})$. Therefore, the general condition to realize a realistic inflationary scenario is
\begin{equation}
\mathcal N_\text{min}=\frac{H_\text{dS}}{A_+}\simeq 60\,,\quad 60\simeq \mathcal N_\text{min}\leq\mathcal N\,.
\end{equation}
The second condition is trivially satisfied by  realistic models of inflation. On the other side,
the first condition fix the curvature expansion rate since, by using (\ref{330}), we obtain
\begin{equation}
R_\text{dS}\simeq R_0\text{e}^{89}\,.
\end{equation}
Now we must involve (\ref{RdSlambda}) in order to fix the boundary term $\lambda(0)$
\begin{equation}
R_{\text{dS}}=R_0\text{e}^{89}\simeq\frac{0.040}{(t_0'\kappa_0^2)\left(\lambda(0)t_0'\right)^{0.77}}\,.
\label{R0}
\end{equation}
Without loss of generality, we can set
\begin{equation}
t_0'=1\,,\label{s1}
\end{equation}
and introduce the Planck mass in $\kappa_0^2$ (\ref{kappa0exp}), namely
\begin{equation}
\kappa_0^2=3.49\times 10^{-55}\text{eV}^{-2}\,.\label{kappaset}
\end{equation}
By taking into account the value of the cosmological constant (see also Ref.~\cite{Barrow}),
\begin{equation}
\Lambda=11.895\times 10^{-67}\text{eV}^2\simeq 10^{-122}M_{\text{Pl}}^2\,,
\label{lambdaset}
\end{equation}
we obtain a realistic ratio $R_\text{dS}/(4\Lambda)=10^{130}$ by setting
\begin{equation}
\lambda(0)=8\times 10^{-16}\,.\label{s4}
\end{equation}
In this case the right hand side of (\ref{condlambda0}) yields $\sim -3.5\times 10^{-4}$ and the condition is well satisfied. Here, we must stress that the model predicts inflation with curvature at least 130 orders of magnitude larger than the curvature of the Universe today. Indeed, if the curvature of inflation is smaller, condition (\ref{condlambda0}) is not fulfilled and the matter/radiation eras with the following late-time acceleration may be not well reproduced, so that a unified description fails.\\
\\
At the end of the early-time acceleration stage some reheating mechanism is involved with the purpose to convert the  inflation energy into standard matter and radiation (e.g., the quark gluon plasma). Since quantum gravity effects then disappear, the gravitational Lagrangian in (\ref{action0}) reads (we still neglect dark energy in $f_\text{DE}(R)$)
\begin{equation}
\mathcal L\simeq \sqrt{-g}\left[\frac{R}{\kappa_0^2}+\frac{0.02}{\lambda(0)}R^2+\frac{1}{\lambda(0)}C^2\right]\,,
\end{equation}
where we have taken into account (\ref{omega0.02}) and (\ref{fullLambda}) with (\ref{parLambda0}). Once again, the square of the Weyl tensor does not contribute to the field equations in FRW space-time and the reheating mechanism coincides with the one for $R^2$-inflation. The first Friedmann-like equation leads to
\begin{equation}
\ddot H-\frac{\dot H^2}{2H}+\frac{\lambda(0)}{0.24\kappa_0^2}H=-3H\dot H\,,
\end{equation}
with the oscillating solution
\begin{equation}
H\simeq\frac{4}{3(t-t_\text{r})}\cos^2\left[
\sqrt{\frac{\lambda(0)}{0.12\kappa_0^2}}\frac{(t-t_\text{dS})}{2}
\right]\,,
\end{equation}
where $t_\text{r}$ is the time at reheating, and, since $<H>\simeq 2/(3(t-t_\text{r}))$, we get a matter-like cosmological evolution. By taking into account that the Hubble parameter tends to vanish at the end of inflation\footnote{In the second reference of \cite{SuperBamba} the critical value of the curvature during reheating for a combined model of $R^2$-inflation with exponential $F(R)$-gravity for dark energy $F(R)=-\beta R_c(1-\text{e}^{-R/R_c})$ has been accurately calculated as $R_{cr}=R_c\log\beta$.}
 ($R_0\ll R_\text{dS}$), and $R\simeq 6\dot H$, we derive for the Ricci scalar
\begin{equation}
R\simeq -\frac{4}{(t-t_\text{r})}\sqrt{\frac{\lambda(0)}{0.12\kappa_0^2}}\sin\left[\sqrt{\frac{\lambda(0)}{0.12\kappa_0^2}}(t-t_\text{r})\right]\,.
\label{Rreheating}
\end{equation}
And using the Lagrangian of a scalar bosonic field\footnote{The production of bosons is favored with respect to that of fermions during reheating~\cite{infrev}.} $\chi$ with mass $m_\chi$ and non-minimally coupled with gravity,
\begin{equation}
\mathcal L_\chi=-\frac{g^{\mu\nu}\partial_\mu\chi\partial_\nu\chi}{2}-\frac{m_\chi^2\chi^2}{2}-\frac{\xi R\chi^2}{2}\,,
\end{equation}
$\xi$ being a coupling constant, one obtains the field equation
\begin{equation}
\Box\chi-m_\chi ^2\chi-\xi R\chi=0\,.
\end{equation}
By decomposing the field $\chi$ into Fourier modes $\chi_k\equiv\chi_k(t)$ with momentum $k$,  on FRW space-time, we get
\begin{equation}
\ddot\chi_k+3H\dot\chi_k+\left(m^2_\chi+\xi R\right)\chi_k=0\,.
\end{equation}
Now we can introduce a conformal time $d\eta=dt/a(t)$, such that
\begin{equation}
\frac{d^2}{d\eta^2}u_k+m_\text{eff}^2 a(t)^2u_k=0\,,\quad u_k= a(t)\chi_k\,,\label{u}
\end{equation}
where the effective mass $m_\text{eff}$ is given by
\begin{equation}
m^2_\text{eff}=\left[m_\chi^2+\left(\xi-\frac{1}{6}\right)R\right]\,.\label{meff}
\end{equation}
Since the Ricci scalar oscillates as in Eq.~(\ref{Rreheating}), the number of massive particles $\chi_k$ changes with time and the reheating mechanism takes place. We should note that, even in the case of minimal coupling with gravity $\xi=0$, the effective mass $m^2_\text{eff}$
still depends on $R$ and we do get reheating. After particle production, when the energy density of radiation and ultrarelativistic matter becomes dominant and the $R^2$-term in the Lagrangian vanishes, due to the condition (\ref{condlambda0}), the usual radiation/matter era can start.

The second part of our work is thus devoted to the study of Friedmann cosmology; in the next section we will introduce modified gravity for the dark energy sector through the function $f_\text{DE}(R)$ in (\ref{action0}). We will see that a complete picture of the matter era and of dark energy de Sitter expansion can be recovered in our model, thus confirming that high-curvature corrections of the model disappear at small curvature, as expected.

\section{Dark energy from exponential gravity}

In Refs.~\cite{nostriexp, altriexp, HS} several versions of viable modified
gravity for the dark energy epoch have been proposed and investigated.
They belong to a class of so-called ``one-step models'', which
produce reasonable description of the dark energy evolution of our Universe today.
They incorporate a vanishing cosmological constant in the flat limit
($R\rightarrow 0$), and exhibit a suitable, constant asymptotic
behavior for large values of the curvature, mimicking in fact the $\Lambda $CDM Model.

In this work we deal with a modified version of exponential gravity~\cite{nostriexp} where,
in order to reproduce the dark energy sector, we introduce the following form of $f_\text{DE}(R)$ in the action (\ref{action0}),
\begin{equation}
f_\text{DE}(R)=-\frac{2\Lambda g(R)(1-\text{e}^{-a R/\Lambda})}{\kappa_0^2}\,,\label{fDE}
\end{equation}
where $a$ is a positive parameter, $\Lambda$ is the cosmological constant whose value has been given in (\ref{lambdaset}), and $\kappa_0^2$ is the mass scale in (\ref{kappa0exp}). Moreover, $g(R)$ is a function of the Ricci scalar which will play an important role at large curvature and will be discussed later. We require that $g(R)=1$ when
$R=0$, in order to recover the Minkowski solution of General Relativity in  flat space-time.

The magnitude of the modified gravity function introduced above can be
estimated as $|f_\text{DE} (R)| \simeq 2\Lambda/\kappa_0^2$. For a comparison with
the higher curvature corrections in (\ref{action0}), we note that at the end of
inflation $R/\kappa^2(t')\simeq R_0/\kappa_0$. For example, by using (\ref{R0})--(\ref{s4}), one finds $|f_\text{DE}(R_0)| /(R_0/\kappa_0^2)\simeq 2.2 \times 10^{-92}$ and we
easily understand that the modification of gravity for the dark energy
sector is completely negligible at the  inflationary era.

Exponential gravity reproduces the cosmological constant at large curvature when
\begin{equation}
f_\text{DE}(R)\simeq -\frac{2\Lambda g(R)}{\kappa_0^2}\,,\quad \Lambda\ll R\,,
\end{equation}
and for this reason we here  set
\begin{equation}
a=\frac{1}{2}\,,
\label{a}
\end{equation}
but, in general, any choice of $0<a\leq 1$ is still reasonable. Therefore, when $g(R)$ remains close to one, we can reproduce the standard $\Lambda$CDM model.

Let us introduce the following definition
\begin{equation}
F(R)\simeq R+\kappa_0^2 f_\text{DE}(R)\,.
\end{equation}
We do not consider the induced quantum corrections of the model because curvature is rather small and quantum gravity effects are not essential.
In what follows, we will study the vacuum de Sitter solution and we will give some general considerations on its behavior at small curvature. In Sections {\bf VI, VII} a detailed analysis on the dark energy expansion in our model will be carried out, and we will consider again the whole form of the Lagrangian discussed in the first section. Thus, it will be clear that quadratic corrections of the theory are negligible in the late Friedmann Universe.

Since modified gravity introduces a new degree of freedom, the following conditions must be unavoidably met
\begin{equation}
|F_R(R)-1|\ll 1\,,\quad 0<F_{RR}(R)\,,\quad\text{when}\quad 4\Lambda<R\,.\label{viableFR}
\end{equation}
The first one is necessary to correctly reproduce the Newton constant, avoiding antigravitational effects during matter, radiation and dark energy eras, while the second condition is necessary to prevent the occurrence of matter instabilities~\cite{Faraoni2, Song, inlation+DE} (see also Ref.~\cite{Cam}). We should note that pure exponential gravity ($g(R)=1$) does satisfy (\ref{viableFR}), but, due to the fact that $R F_{RR}(R)\simeq 0^+$ for large values of $R$, some singular solution may emerge in the theory~\cite{oscillation1}. In Ref.~\cite{oscillation2} where the problem has been studied in detail, the singularities are avoided with the introduction of a $R^{1/3}$-term in the gravitational action, while here the ``curing'' term is included in the dark energy function $f_\text{DE}(R)$ through $g(R)$. We propose the following form of $g(R)$,
\begin{equation}
g(R)=\left[1-b\left(\frac{R}{4\Lambda}\right)\log\left[\frac{R}{4\Lambda}\right]\right]\,,\quad
0<b\,,\label{gR}
\end{equation}
$b$ being a constant positive parameter. We will better understand the motivations of this choice in the next section. Here, we analyze the behavior of the model at small curvature.
The field equations of the theory can be written as
\begin{equation}
3\Box F_R(R)+R F_R(R)-2F(R)=\kappa_0^2 T\,,\label{tracefield}
\end{equation}
where $T$ is the trace of the stress-energy tensor associated with the matter Lagrangian.
We immediately observe that,
since $g(R)=1$ and $f_{\text{DE}}(R)\simeq -2\Lambda$ when $R=4\Lambda$, the model admits the (vacuum) de Sitter solution and leads to the accelerated expansion of our Universe today.
By considering a perturbation $\delta R\equiv \delta R(t)$, $|\delta R/R|\ll 1$, around the de Sitter solution, we derive from (\ref{tracefield}) in vacuum
\begin{equation}
\left(\ddot\delta R+3H\dot \delta R - m_\text{eff}^2\delta R\right)\simeq 0\,,\quad m_\text{eff}^2=\frac{1}{3}
\left(\frac{F_R(R)}{F_{RR}(R)}-R
\right)\,,
\end{equation}
and the solution is stable when the effective mass $m_\text{eff}^2$ is positive, namely ~\cite{proc},
\begin{equation}
1<\frac{F_R(R)}{R F_{RR}(R)}\,.
\end{equation}
Since for our model $F_R(R)\simeq 1$ and $F_{RR}(R)\simeq b/(2R)$ when $R=4\Lambda$, it is easy to see that, if $0<b$, the de Sitter solution is stable and it is moreover a final attractor of the system, when in the expanding Universe the contents of matter and radiation vanish (namely, in the asymptotic future).\\
\\
Before passing to the next section where the features of the function $g(R)$ will be investigated in the context of the radiation/matter evolution ($T\neq 0$ in (\ref{tracefield})), we here conclude with some observations about the parameter $b$ of $g(R)$. Since we would like to maintain the same behavior of the  $\Lambda$CDM model at late time, we must set the parameter $b$ according to the condition
\begin{equation}
b\ll \left[\left(\frac{R}{4\Lambda}\right)\log\left[\frac{R}{4\Lambda}\right]\right]^{-1}\,,\quad
4\Lambda\leq R\ll R_0\,,
\label{bcond}
\end{equation}
where we recall that $R_0$ is the curvature of the Universe at the end of inflation.
A reasonable choice is
\begin{equation}
b=10^{-5}.\label{bbbb}
\end{equation}
In this case the condition (\ref{bcond}) is satisfied also at still higher curvatures (matter era), up to the value $R\simeq 4\Lambda\times 10^{4}$. For larger values of the curvature,  matter and radiation are highly dominant, but conditions (\ref{viableFR}) are still required and are well satisfied when $4\Lambda\leq R\ll R_0$. In this case,
\begin{equation}
\left(F_R(R)-1\right)\simeq \frac{b}{2}\left[1+\log\left[\frac{R}{4\Lambda}\right]\right]\,,\quad
F_{RR}(R)\simeq\frac{b}{2R}\,.
\end{equation}
For example, if $R=R_0$, where $R_0$ is given by (\ref{R0}) with (\ref{s1})--(\ref{s4}), namely $R_0\simeq 2\times 10^{91}\times 4\Lambda$, we get $(F_R(R)-1)\simeq 10^{-3}$ and the model is not affected by antigravity. Finally, when $b$ is positive the second condition in (\ref{viableFR}) is automatically satisfied.

\section{Radiation/matter evolution}

In this section we study the evolution of the model during the radiation and matter eras. When matter or radiation are dominant, an oscillating behavior of the dark energy may affect the solution and could bring to the appearance of unphysical singularities. This is a well-known problem of modified gravity for dark energy, strictly related to the fact that $R F_{RR}(R)\ll 1$ but $F_{RR}(R)\neq 0$, at large curvature, and requires a careful investigation.

If one introduces the variable
\begin{equation}
y_H \equiv\frac{\rho_{\mathrm{DE}}}{\rho_{\mathrm{m}(0)}}\,,
\label{y}
\end{equation}
where $\rho_\text{DE}$ is the dark energy density at some time and $\rho_{\text{m}(0)}$ the matter energy density\footnote{The matter energy density includes both cold and baryonic matter.} of the Universe today, the implicit form of the first Friedmann-like equation reads
\begin{equation}
y_H (z)=\frac{H(z)^2}{m^2}-(z+1)^3-\chi
(z+1)^{4}\,,
\label{y}
\end{equation}
where $y\equiv y_H(z)$ and $H\equiv H(z)$ have been expressed as functions of the redshift normalized to one at present, namely $z=\left[1/a(t)-1\right]$, $\chi$ is referred to as the energy density of radiation today $\rho_{\mathrm{r}(0)}$, namely $\chi\equiv\rho_{\mathrm{r}(0)}/\rho_{\mathrm{m}(0)}$,
and $m^2$ is the mass scale associated with the Planck mass; it is given by
\begin{equation*}
m^2\equiv\frac{\kappa_0^2\rho_{\mathrm{m}(0)}}{6}\,,
\end{equation*}
with $\kappa_0^2$ as in (\ref{kappaset}).
Cosmological data lead to
\begin{equation}
m^2\simeq 1.82 \times
10^{-67}\text{eV}^2\,,\quad \chi\simeq 3.1 \times
10^{-4}\,.
\end{equation}
The FRW field equations of the model supplied by the usual conservation laws of matter and radiation can be recast as~\cite{SuperBamba}
\begin{equation}
\frac{d^2 y_H(z)}{d z^2}+J_1\frac{d y_H(z)}{d z}+J_2
y_H(z)+J_3=0\,,
\label{superEq}
\end{equation}
with
\begin{eqnarray}
J_1 \Eqn{=} \frac{1}{(z+1)}\left[-3-\frac{1}{y_H+(z+1)^{3}+\chi (z+1)^{4}}\frac{1-F_R(R)}{6m^2
F_{RR}(R)}\right]\,, \nonumber
\\
J_2 \Eqn{=}
\frac{1}{(z+1)^2}\left[\frac{1}{y_H+(z+1)^{3}+\chi (z+1)^{4}}\frac{2-F_R(R)}{3m^2 F_{RR}(R)}\right]\,, \nonumber
\\
J_3 \Eqn{=}
-3 (z+1)
\nonumber \\
&&
-\frac{(1-F_R(R))((z+1)^{3}+2\chi (z+1)^{4})
+(R-F(R))/(3m^2)}{(z+1)^2(y_H+(z+1)^{3}+\chi
(z+1)^{4})}\frac{1}{6m^2
F_{RR}(R)}\,.\label{J3}
\end{eqnarray}
Now, $F(R)$ corresponds to the whole gravitational Lagrangian of the model in the FRW space-time, namely
\begin{equation}
F(R)=\kappa_0^2\left[\frac{R}{\kappa^2(t')}-\tilde\Lambda(t')-\frac{\omega(t')}{3\lambda(t')}R^2+ f_{\text{DE}}(R)\right]\,,\label{FR}
\end{equation}
where $t'$ is given by (\ref{t'}) and the Ricci scalar (\ref{Ricci}) is derived from (\ref{y}), as
\begin{equation}
R=3 m^2 \left[4y_H(z)-(z+1)\frac{d y_H(z)}{d z}+(z+1)^{3}\right]\,.
\label{RRR}
\end{equation}
At late times ($z\ll 1$), when dark energy is dominant and the contribution of matter $\sim (z+1)^3$ is negligible in (\ref{RRR}), the solution of (\ref{superEq})--(\ref{J3}) is
\begin{equation}
y_H\simeq\frac{\Lambda}{3m^2}+y_0\text{Exp}\left[\pm i\sqrt{\frac{1}{\Lambda F_{RR}(4\Lambda)}-\frac{25}{4}}\log[z+1]
\right]\,,
\end{equation}
with $y_0$ an integration constant. For our model, as $\Lambda F_{RR}(4\Lambda)\ll1 $, the argument of the square root is positive. For example, by setting $a\,,b$ as in (\ref{a}) and (\ref{bbbb}), respectively, given $\Lambda$ as in (\ref{lambdaset}), and using the parameterization in (\ref{fullLambda})--(\ref{parLambda0}) and (\ref{R0})--(\ref{s4}) for the higher curvature corrections of the model, we get $[1/(\Lambda F_{RR}(4\Lambda))-25/4]\simeq 366.37$. As a consequence, the dark energy develops an oscillatory behavior~\cite{staroomega}. In our example, we derive that the frequency of oscillations $\nu$ with respect to $\log[z+1]$ is $\nu\simeq \sqrt{366.37}/(2\pi)\simeq 3.04$, confirming the stability of the de Sitter solution.

At large curvature, the oscillation frequency of dark energy can diverge. If we assume that the contribution of dark energy in (\ref{RRR}) is negligible ($0\ll z$), after the expansion of (\ref{superEq})--(\ref{J3}) with respect to $y_H/(z+1)^3\ll 1$, we find the following solution in the vicinity of a given redshift $z$~\cite{oscillation1, oscillation2},
\begin{equation}
y_\text{H}(z+\delta z)\simeq \frac{\Lambda}{3m^2}+y_0\text{Exp}\left[\pm i\nu \delta z\right]\,,
\end{equation}
where $|\delta z/z|\ll 1$, $y_0$ is an integration constant, and the oscillation frequency of dark energy $\nu$ reads
\begin{equation}
\nu\simeq \frac{1}{2\pi\sqrt{R F_{RR}(R)}(z+1)}\,.
\end{equation}
Note that the  dark energy oscillations are amplified in the related dark energy equation of state (EoS) parameter, as
\begin{equation}
\omega_\text{DE}(z)=-1+\frac{1}{3}(z+1)\frac{d y_H(z)}{dz}\,,\label{omegaDE}
\end{equation}
which oscillates around the line of the phantom divide.

We see that the model can correctly reproduce the quasi constant dark energy amount $y_H(z)\simeq \Lambda/(3m^2)$ with EoS parameter $\omega_\text{DE}\simeq -1$, but if $0<R F_{RR}(R)\ll 1$ the dark energy density oscillates with high frequency and some singularities may emerge in the solution, rendering it unphysical. This is the case of pure exponential gravity, where at large redshift the dark energy frequency diverges and  singularities appear.

As we have recalled above, in Ref.~\cite{oscillation2} an additional term proportional to $-R^{1/3}$ has been added to the model to solve the problem. As a consequence, one has $\nu\propto R^{1/3}/(z+1)$ and at high redshift, at least during the matter era with $R\simeq 3m^2(z+1)^3$, the oscillation frequency of the dark energy turns out to be constant, stabilizing the theory. In our case, thanks to the introduction of $g(R)$ as in (\ref{gR}), we obtain
\begin{equation}
\nu\simeq\frac{\sqrt{2/b}}{2\pi(z+1)}\,.
\label{nu}
\end{equation}
This result is independent of the on-shell form of the Ricci scalar (radiation/matter solution) and it shows that the frequency of the dark energy oscillations decreases back into the past, thus avoiding any kind of singularity. In the next section we will provide a numerical simulation of the dark energy evolution in our model. We will confirm our results, obtained in a semi-analytic way, and will prove that the model is able to reproduce the late-time expansion of the Universe in accordance with the astronomical data.

\section{A numerical simulation}

In this section we will proceed with a numerical test of our model at late times. We will use the system (\ref{superEq})--(\ref{J3}) and the whole form of the Lagrangian in (\ref{FR}). The Planck mass and cosmological constant are fixed as in (\ref{kappaset})--(\ref{lambdaset}). The forms of $\lambda(t')$ and $\omega(t')$ are given by (\ref{lambda}) and (\ref{omega0.02}), respectively, with $\beta_2$  in (\ref{betabeta}). The form of $\tilde\Lambda(t')$ is given by (\ref{fullLambda}) with (\ref{parLambda0}) and we can drop it down, as explained.
For $\lambda_0$,  the boundary term of inflation, we have used the value in (\ref{s4}). The parameter $t'$ is defined by (\ref{t'}), with (\ref{R0}) and (\ref{s1}). Remember that, with such kind of parametrization, we get a realistic inflationary scenario at high curvature. Finally, the dark energy function $f_\text{DE}(R)$ is given by (\ref{fDE})--(\ref{gR}), where for the parameters $a$ and $b$ we may choose the values in (\ref{a}) and (\ref{bbbb}).

We have performed a numerical simulation\footnote{Mathematica 8 \copyright.} for $-1<z<z_\text{max}$ with $z_\text{max}=10$. Thus, we need the boundary conditions of the system at $z=z_\text{max}$, namely
$y_H(z_\text{max})$ and $dy_H(z_\text{max})/dz$. They can be derived from the explicit form of $\rho_\text{DE}$ in (\ref{y}) for $F(R)$-gravity, namely
\begin{equation}
\rho_\text{DE}=
\frac{1}{\kappa_0^2 F_R(R)}\left[
(R F_R(R)-F(R))-6H\dot F_R(R)
\right]\,.
\end{equation}
In our case, when $\Lambda\ll R\ll R_0$, we get
\begin{equation}
y_H(z)\simeq\left(\frac{\Lambda}{3m^2}\right)\left(g(R)+6H g_{RR}(R)\dot R\right]\,,
\end{equation}
where $R\equiv R(z)$ and $H\equiv H(z)$ such that $\dot R= -H(z+1)(d R/dz)$.
At large redshift $z=z_\text{max}$,
by assuming $y_H$ to be negligible in (\ref{RRR}), we can take $R=3m^2(z+1)^3$ and  $H=m (z+1)^{3/2}$. Here, we are avoiding the radiation contribution also, namely we are in the case $\chi(z+1)\ll 1$.
As a consequence, the boundary conditions of the system are given by
\begin{eqnarray}
y_H(z_\text{max})&=&\left(\frac{\Lambda}{3m^2}\right)\left[g(R_\text{max})-54 m^4(z_\text{max}+1)^6 g_{RR}(R_\text{max}) \right]\,,\nonumber\\
\frac{d y_H}{d z}(z_\text{max})&=&
3\Lambda(z+1)^2\left[
g_R(R_\text{max})-6 R_\text{max}^2 g_{RRR}(R_\text{max})-12 R_\text{max} g_{RR}(R_\text{max})
\right]\,,
\end{eqnarray}
with
\begin{equation}
R_\text{max}=3m^2(z_\text{max}+1)^3\,.
\end{equation}
We start our simulation at $z_\text{max}=10$, so that
\begin{equation}
y_H(z_\text{max})=2.1818\,,\quad
\frac{d y_H}{d z}(z_\text{max})=-0.0000260653\,,\quad z_\text{max}=10\,.
\end{equation}
Note that $\chi(z_\text{max}+1)\simeq 0.00341\ll 1$, and we effectively are in a matter dominated Universe. We should also stress that the chosen redshift range $-1<z<10$ includes
a very important part of the history of our Universe. Just recall that  the age of the first observed galaxies corresponds to a redshift $z \simeq 6$, even if the existence of older galaxies cannot be excluded.

It is interesting to compare the values above with the corresponding ones for the $\Lambda$CDM model, namely
\begin{equation}
y_H=\left(\frac{\Lambda}{3m^2}\right)=2.17857\,,\quad \frac{d y_H}{d z}=0\,.
\end{equation}
Thus, we see that our model remains extremely close to the $\Lambda$CDM model at high redshift.

\begin{figure}
\begin{center}
\includegraphics[angle=0, width=0.65\textwidth]{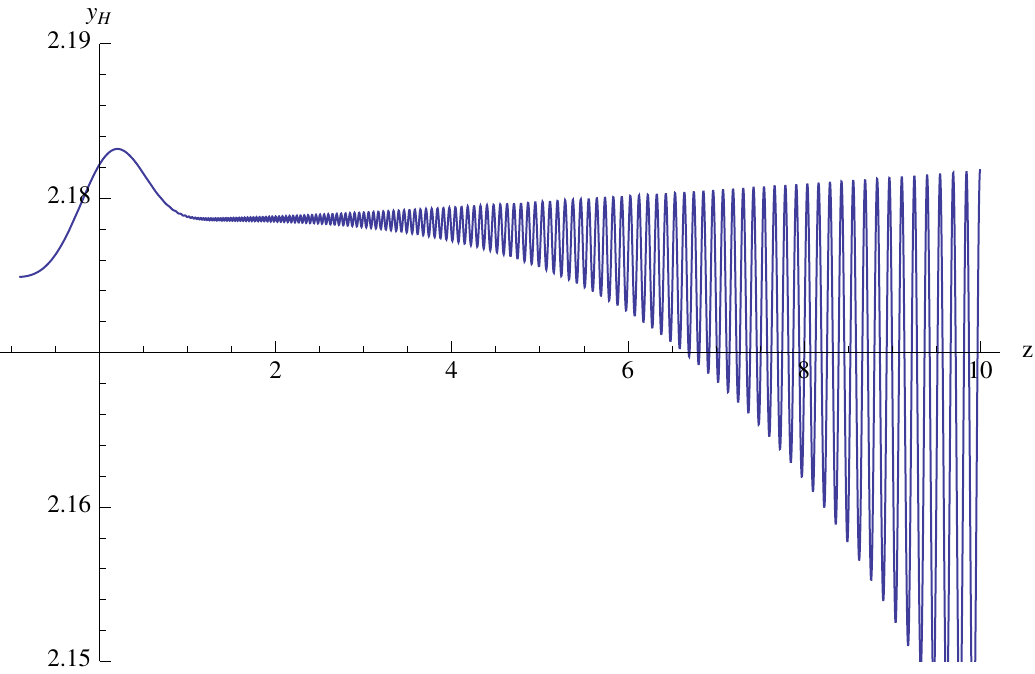}
\end{center}
\caption{Plot of $y_H(z)$ for $-1<z<10$.\label{Fig1}}
\end{figure}
\begin{figure}
\begin{center}
\includegraphics[angle=0, width=0.65\textwidth]{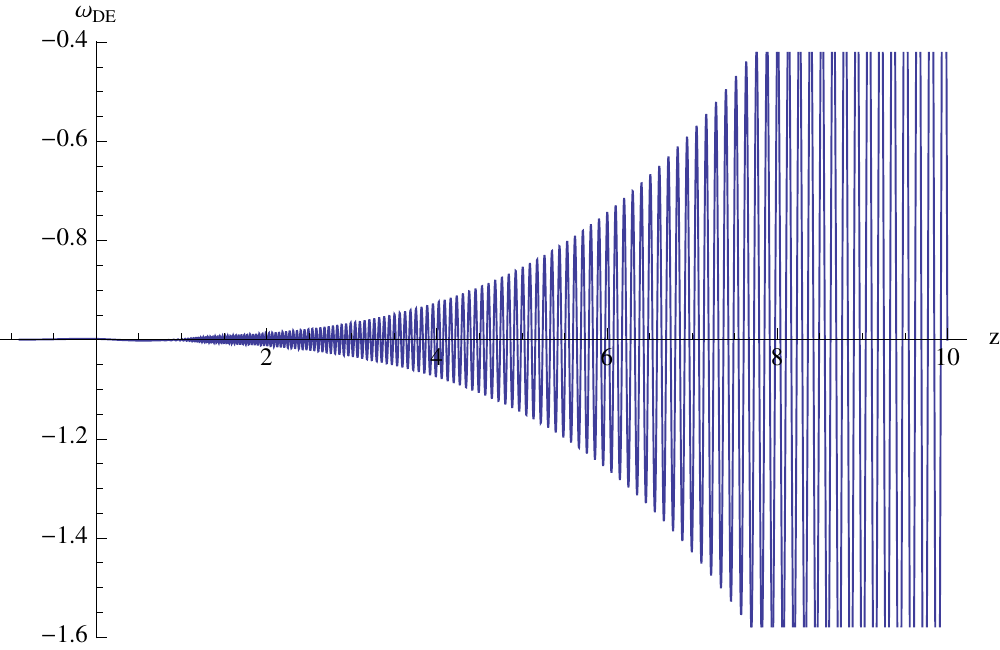}
\end{center}
\caption{Plot of $\omega_\text{DE}(z)$ for $-1<z<10$.\label{Fig2}}
\end{figure}

In Fig.~(\ref{Fig1}) and Fig.~(\ref{Fig2}) we plot  $y_H(z)$ and $\omega_\text{DE}(z)$ (\ref{omegaDE}), respectively, for $-1<z<10$. At high redshift the oscillations of the dark energy density are amplified in its EoS parameter but their frequency does not diverge and, as we will better see later, it tends to decrease back into the past, when the curvature increases.

\begin{figure}
\begin{center}
\includegraphics[angle=0, width=0.65\textwidth]{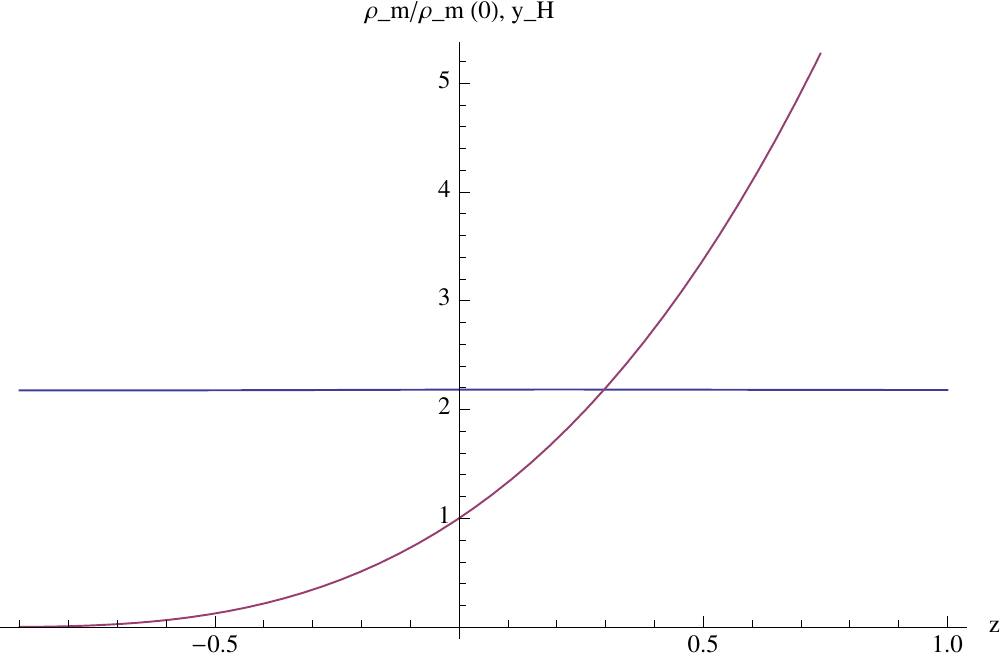}
\end{center}
\caption{Plot of $\rho_\text{m}(z)/\rho_\text{m(0)}$ and $y_H(z)$ for $-1<z<1$.\label{comparison}}
\end{figure}

In Fig.~(\ref{comparison}) we compare the plots of $y_H(z)$ and  $\rho_\text{m}(z)/\rho_{\text{m}(0)}=(z+1)^3$, $\rho_\text{m}(z)$ being the matter energy density at a given redshift $z$, for $-1<z<1$. We clearly see that at the cosmological level $y_H(z)$ is almost a constant, while the energy density of matter decreases with the redshift. The dark energy epoch starts at $z<0.4$, in full agreement with the $\Lambda$CDM description.

\begin{figure}
\subfigure[]{\includegraphics[width=0.4\textwidth]{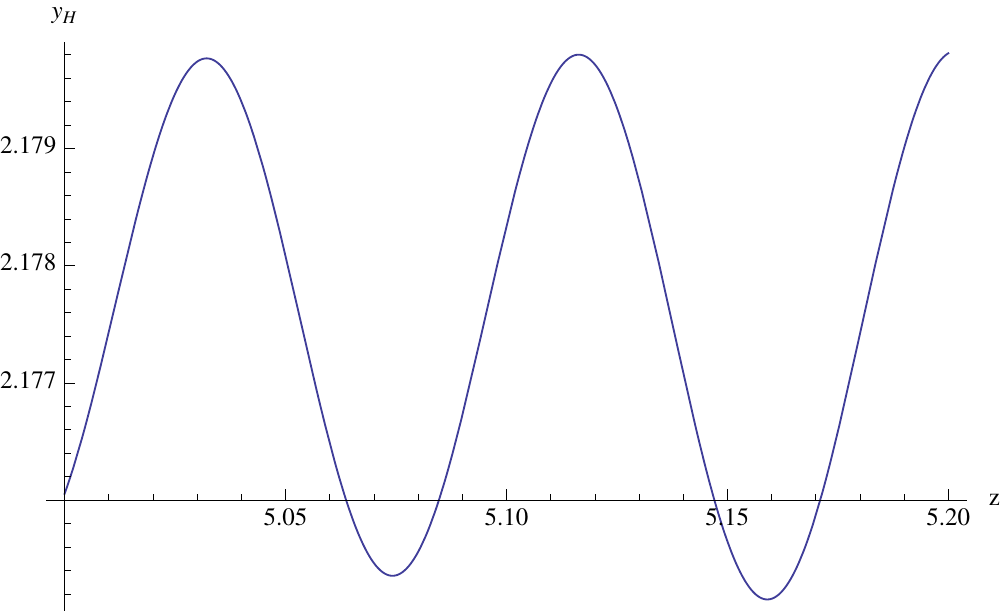}}
\qquad
\subfigure[]{\includegraphics[width=0.4\textwidth]{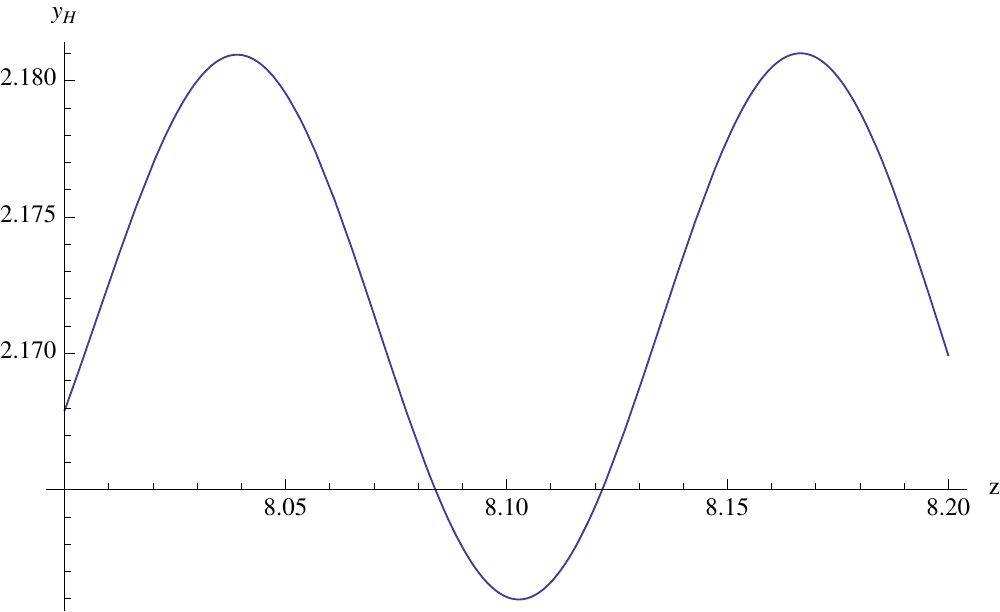}}
\caption{
Plot of $y_H(z)$ for $5<z<5.2$ (a) and $8<z<8.2$ (b).
\label{34}}
\end{figure}

To show how the oscillation frequency of dark energy decreases with increasing redshift (i.e., the curvature in the Friedmann expanding Universe), in Fig.~(\ref{34}) we plot $y_H(z)$ for the intervals $5<z<5.2$ (on the left) and $8<z<8.2$ (on the right), respectively. The numerical simulation confirms the result in (\ref{nu}) which implies that, given a small redshift interval $\delta z$ in the vicinity of the redshift value $z$, the number of dark energy oscillations $n(z)$ turns out to be
\begin{equation}
n(z)=\sqrt{\frac{2}{b}}\frac{\delta z}{2\pi(z+1)}\,.
\end{equation}
Thus, by setting $\delta z=0.02$, we obtain $n(z=5)\simeq 2.37$ and $n(z=8)\simeq 1.58$ as in the plots.

At small redshift the system is dominated by dark energy.
We confirm that the de Sitter solution is a final attractor of the system. If we introduce the parameter $\Omega_\text{DE}(z)$ as the ratio between the dark energy density $\rho_\text{DE}$ and the effective energy density $\rho_\text{eff}$ of our FRW Universe,
\begin{equation}
\Omega_\text{DE}(z)\equiv\frac{\rho_\text{DE}}{\rho_\text{eff}}=\frac{y_H(z)}{y_H(z)+(z+1)^3+\chi(z+1)^4}\,,
\label{OMEGADE}
\end{equation}
by extrapolating $y_H(z)$ at the current redshift $z=0$, from (\ref{OMEGADE}) and (\ref{omegaDE}), we have
\begin{equation}
\Omega_\text{DE}(z=0)=0.685683\,,\quad\omega_\text{DE}(z=0)=-0.998561\,,
\end{equation}
in agreement with the most recent cosmological data analysis~\cite{Planck}, which give $\Omega_\text{DE}(z=0)=0.685\pm0.013$ and $\omega_\text{DE}(z=0)=-1.006\pm0.045$.

In conclusion, we have proven that the model can correctly reproduce the late-time acceleration of the Universe, being also perfectly compatible with inflation at high curvature.

\section{Conclusion}

We have obtained in this work a quite natural unified description of the early- and late-time accelerated expansion of our Universe at two completely different energy scales. To achieve this nice result, we have considered a gravitational Lagrangian with quadratic corrections for inflation and a contribution from phenomenological exponential $F(R)$-modified gravity for the dark energy sector.

As conveniently explained in the paper, these high-curvature corrections to  Einstein's theory are well motivated by quantum gravity. And, given the fact that the Hilbert-Einstein action for GR is not renormalizable, it is also compulsory to consider an effective improved action with an ultraviolet completion at high-energy scale, which might eventually lead to a renormalizable theory. As a starting step in this direction, in this paper we have used a rather simple model for renormalizable quadratic gravity but we have already found that unstable early-time inflation arises in our model at high curvature. We have also shown that our model for inflation can be naturally extended to include in the action the Gauss-Bonnet and $\Box R$ terms, with respective running coupling constants.\footnote{These terms appear in more fundamental formulations, as string theory and others.} The new set of RG equations for the coupling constants, corresponding to this case, have been derived.

We have obtained an early-time inflationary stage consistent with the cosmological data and yielding an $e$-fold number, which is large enough in order to lead to the necessary thermalization of our observable Universe, and also a spectral index and tensor-to-scalar ratio that fit very well the most accurate Planck results. Moreover, at the end of inflation, when the quantum effects disappear, our model evolves into the usual $R^2$ correction to GR, and provides a perfectly appropriate reheating mechanism, as well.

Here, we would like to point out that the presence of the square of
the Weyl tensor in the gravitational Lagrangian, which is fundamental in
order to deal with a renormalizable theory, may only modify the study of cosmological perturbations.
 In a pure conformally invariant theory, the
contribution of the Weyl tensor vanishes in the primordial scalar power
spectrum and leads to constant terms in the spectra of vector and
tensor perturbations. It is not clear what happens when one considers a
more involved theory. In our case, where the Weyl tensor is coupled with
the Ricci scalar through the relative running coupling constant, the change
of the curvature rate of the model during the exit from inflation may be
modified in order to correctly predict the cosmological data. On the other
hand, the lower bound of the e-folds number, which is found in the FRW
background and is independent on the Weyl term, is still fixed at $\mathcal
N_{\text{min}} \simeq 60$. For these reasons, we may argue that the spectra of the
perturbations left at the end of inflation remain realistic, but this point
deserves for sure a further investigation.

Exponential modified gravity for dark energy offers an accurate description of the current cosmic accelerated expansion. However, during the radiation and matter domination eras, singularities could emerge from the theory, thus rendering its solutions nonphysical. The problem stems from the fact that in modified gravity a new degree of freedom appears, which leads to a dark energy oscillatory behavior whose frequency may diverge at large curvature. For this reason, we have introduced a logarithmic correction to the cosmological constant parameter of the exponential $F(R)$-modified gravity function appearing in the action. 
Such correction is qualitatively similar to the ones induced by quantum
corrections at high curvature, but it works at the current scale
 and must be interpreted as a phenomenological modified gravity term.
Thanks to our curing term, the theory turns out to be free from singularities and, thus, fitting simulations at large redshift can be readily carried out. The results of our semi-analytic analysis are  well confirmed by these numerical simulations. We have also shown that the different stages of the universe evolution, namely radiation/matter and dark energy domination, do take place in our model after the inflation epoch, similarly to the $\Lambda$CDM standard model. Also, we have shown that, in our model, the values of the basic cosmological parameters resulting from the latest and most accurate analysis of the data obtained with the Planck satellite, which provide the most reliable constrains on the nature of dark energy, can be faithfully recovered.

\medskip

\noindent {\bf Acknowledgements}.  EE and SDO are partially supported by CSIC, I-LINK1019 Project, by MINECO (Spain), Projects FIS2013-44881-P and FIS2016-76363-P,  and by the CPAN Consolider Ingenio Project.

\end{document}